\magnification=\magstep1 
\baselineskip=14 pt
\hsize=5 in
\vsize=7.3 in
\pageno=1
\line{\hfill hep-th/9601103}
\vskip 1,8 cm
\centerline{\bf THERMAL CONDITIONS FOR SCALAR BOSONS}
\centerline{\bf IN A CURVED SPACE TIME}
\vskip 1,8 cm 
\centerline{\bf Carlos E. Laciana}
\centerline{\sl Instituto de Astronom\'{\i}a y F\'{\i}sica del Espacio} 
\centerline{\sl Casilla de Correo 67 - Sucursal 28, 1428 Buenos Aires, 
Argentina}
\centerline{\sl E-mail: laciana@iafe.uba.ar}
\vskip 1,8 cm
{\bf Abstract}
\vskip 0,2cm 
 The conditions that allow us to consider the 
vacuum expectation value  of the energy-momentum
 tensor as a statistical average, at some 
particular temperature, are given. When the mean value of created 
particles is stationary, a planckian distribution for the field modes is 
obtained. In the massless approximation, the temperature dependence is as that  
corresponding to a radiation dominated Friedmann-like model.   
\vskip 2,5 cm
\vfill\eject
1.{\bf Introduction}

 In the semiclassical approximation the vacuum expectation value (VEV) 
of the energy-momentum tensor (EMT) can be used as a 
source of the Einstein equations, in the form:
$$G_{\mu \nu} = -8\pi G<T_{\mu \nu}>_{reg.}\eqno (1)$$

 This approximation  is  useful  as  an  asymptotic  value  of more complete
theories where the gravitational field is also quantized,  
when the time  is  larger  than  the  Planck time.  The mean value of $T_{\mu
\nu}$ can be obtained using the expression: 
$$<T_{\mu \nu}>={{<out,0|{\hat T}_{\mu \nu}|0,in>}\over  {<out,0|0,in>}}\eqno
(2)$$     
\vskip 0,3cm
which was  introduced  in  ref.[1].    We  can  relate  eq.   (2) with 
the VEV with respect to some particular ``in" or ``out" 
vacua (see ref.[2]), i.e.:
$$<T_{\mu \nu}>=<in,0|{\hat T}_{\mu  \nu}|0,in>  + {i\sum_{i,j}}{\Lambda}_
{ij} T_{\mu \nu} ({\!{\phi}^{*}}_{in,i},{\!{\phi}^{*}}_{in,j})$$
$$=<out,0|{{\hat T}_{\mu \nu}}|0,out> + {i\sum_{i,j}}V_{ij} 
T_{\mu \nu}({\! {\phi}^{*}}_{out,i}, {\! {\phi}^{*}}_{out,j})\eqno (3)$$
\vskip 0,3cm
where  $\{{\phi}_{in}\}\bigcup \{{\!{\phi}^{*}}_{in}\}$ 
and $\{{\phi}_{out}\}\bigcup \{{\!{\phi}^{*}}_{out}\}$ are 
the basis of solutions defined for the Cauchy surface labeled by ``in" 
and ``out" 
respectively. ${\Lambda}_{ij}$ and $V_{ij}$ are functions of the Bogoliubov 
coefficients that determine the transformation between the two vacua. It is 
also useful  to  take  into  account  that  the  expectation values can be
expanded  in  terms   of  the vacuum  polarization,  which  contains 
 the  local
infinities, that are  
removed by the usual regularization methods (see ref.[3]) independently of the 
vacuum definition used
,  and a term which  is  function  of  the  created  particles,  due  to  the
interaction of the field with the curved geometry.  This last term is of course 
null in flat space time.  But in curved space it is in general infinite when 
it is calculated perturbatively.  For instance
in  the  particular  case of fields minimally  coupled  to  gravity,  the
infinity  coming from the particle creation term, cannot  be removed  with  the
usual regularization methods  for an arbitrary vacuum definition (see 
ref.[4]). However it is easy to see (ref. [5]) that when particle creation 
has a planckian distribution, which is lost when perturbative 
expansions are performed, the contribution to the EMT is finite and moreover 
is coincident with the standard cosmology which uses  
perfect fluid classical sources  in the 
Einstein equations. A question that we can  ask is;  
if it is reasonable  to get a planckian distribution for 
the mean value of created
particles. The answer is  afirmative  because is obvious the fact that the 
actual  background radiation of the Universe has a black body distribution with 
temperature approximately equal to $3^{o} K$ (see ref.  [6]).  In another case, 
for  example in  the  black  holes  studied  by  Hawking  [7],  a  planckian
distribution is predicted.   Also  the  distribution  seen  by  an accelerated
observer  (Rindler  observer)  is  planckian    like,    therefore    by  the
generalization of the equivalence principle we hope that for some Bogoliubov 
transformations we get a thermal spectrum.   This result is shown in the present
work for a scalar field minimally coupled to a Robertson-Walker metric. The 
expression obtained for the VEV of the EMT  is compared 
with the one coming from 
the statistical  average  of  $T_{\mu \nu}$ at a given temperature, 
which is obtained using Thermo Field Dynamics (TFD) theory [9].  
In a previous work [10] it was proved that a field in a 
curved geometry presents  a  form similar to that of 
the field in a thermal bath.  In
TFD the interaction with the bath is half-filled  
by the so called ``tilde" modes, 
which is more natural  that  the interaction between the classical background
and the quantum field. In the present work from, the comparison mentioned 
before, sufficient conditions on the Bogoliubov transformation are obtained, in 
order to get the same functional form for VEV of the EMT and the statistical 
average obtained from TFD.
\vskip 1,2 cm
   
2. {\bf Time dependent Bogoliubov transformation}
  
 In a pioneer work [11] Parker developed a consistent method to obtain the
\vskip 0,1cm 
annhilation-creation operators at each time in curved space. The operators 
defined at different Cauchy surfaces are related by time dependent  
Bogoliubov transformations. Parker found in ref.[11] the differential 
equation  satisfied by  the Bogoliubov coefficients.  This equation is other
form to write the field equation. In order to resolve that equation it is 
necessary to give some initial conditions. The Bogoliubov coefficients give 
us the mean value of created particles between an initial time ``$t_{0}$" 
and a present time ``$t$".

 In order  to  make   the paper more comprehensive, in this section we repeat in
brief the formulation given by Parker in ref. [11]. 

 The action for  a minimally coupled scalar matter field is: 
$$S={1\over 2}\int {\sqrt {-g}} {d^4}x 
({{{\partial}_{\mu}}\varphi}{{{\partial}^{\mu}}{\varphi}} -
{m^2}{{\varphi}^2})\eqno (4)$$
\vskip 0,2 cm
and the metric is the spatially flat Robertson-Walker, given by 
$${ds^2} = {dt^2} - {a(t)^2}({dx^2} + {dy^2} + {dz^2})\eqno (5)$$ 
\vskip 0,2 cm
Therefore the variation of the action gives the field equation 
$$({{\bigtriangledown}_{\mu} {\partial}^{\mu}}  +  {m^2}){\varphi}  =  0\eqno
(6)$$
\vskip 0,2 cm
(with $\mu = 0,1,2,3$). 
\vskip 0,2 cm
 The metric given by eq.(5) represents an unbounded space-time, therefore 
in a Fourier  representation of $\varphi$  integrals appear, in order to make
 the calculation easier, we can use a discretization in the same form as in  
ref.[11], i.e. we can introduce the periodic boundary condition 
$\varphi ({\bf x} + {\bf n} L, t) = \varphi ({\bf x}, t)$, where $\bf n$ is 
a vector with integer Cartesian components and $L$ a longitude which goes to 
infinity at the end of the calculation (really in our case it is not necessary 
because the results used do not depend on L). Then we can introduce the set of 
functions $\{{\phi _{\bf k}}(x)\}\cup \{{{\! \phi ^{\ast}}_{\bf k}}(x)\}$ 
defined by 
$${{\phi _{\bf k}}(x)} = {1\over {(L a(t))^{3/2} {\sqrt {2W}}}} \exp {i(
{{\bf k}{\bf x}} - {\int _{t_{0}}}^{t} W(k, {t^{\prime}}) dt^{\prime})}
\eqno (7)$$ 
\vskip 0,2 cm
with $W$ an arbitrary real function of $k = \mid {\bf k} \mid$ and $t$. 
The field can be expanded in the form 
$${\varphi ({\bf x}, t)} = {\sum _{\bf k}}[{a_{\bf k}}(t) 
{\phi _{\bf k}}(x) + {{a^{\dagger}}_{\bf k}}(t) 
{{\! \phi ^{\ast}}_{\bf k}}(x)]\eqno (8)$$ 
\vskip 0,2 cm
where the operators $a_{\bf k}(t)$ and $a_{\bf k^{\prime}}(t)$ are time  
dependent and satisfy the commutation relations 
\vskip 0,2 cm 
$$[a_{\bf k}(t), a_{\bf k^{\prime}}(t)] = 0,\ \ \ 
[{{\! a^{\dagger}}_{\bf k}}(t), 
{{\! a^{\dagger}}_{\bf k^{\prime}}}(t)] = 0,\ \ \ 
[{a_{\bf k}}(t), {{\! a^{\dagger}}_{\bf k^{\prime}}}(t)] = 
\delta _{{\bf k}, {{\bf k}^{\prime}}}\eqno (9)$$
\vskip 0,2cm
The operators act on the vacuum $|0,t>$ in the form 
\vskip 0,2 cm
$${a_{\bf k}}(t)|0,t>=0,\  \  \  {{\!  a^{\dagger}}_{\bf k}}(t)|0,t>=|{1_{\bf
k}},t>$$
\vskip 0,2 cm
Moreover we define the operators
$${A_{\bf k}}:= {a_{\bf k}}(t=t_{1})\ \ \ with\ \ t\ge t_{1}\ge t_{0}$$
$${A_{\bf k}}^{\dagger}:={{a_{\bf k}}^{\dagger}}(t=t_{1})\eqno (10)$$ 
Also we define the vacuum $|0>:=|0,t=t_{1}>$ on which the operators 
defined by eq.(10) act. These operators are related with the time dependent 
ones by the following Bogoliubov transformation 
\vskip 0,2 cm
 
$${a_{\bf k}}(t) = {{\alpha ({\bf k},t)}^{\ast}}A_{\bf k}  + 
\beta ({\bf k},t) {A^{\dagger}}_{-{\bf k}}\eqno (11a)$$
\vskip 0,2 cm
$${a_{-{\bf k}}}^{\dagger}(t) = \alpha ({\bf k},t)
{A^{\dagger}}_{-{\bf k}}  + {{\beta ({\bf k},t)}^{\ast}}A_{\bf k} \eqno (11b)$$
\vskip 0,2 cm
clearly we use the conditions 
$$\alpha (k, t_{1}) = 1, \ \ \ \ \beta (k, t_{1}) =  0\eqno (12)$$             

 We can rewrite the field $\varphi ({\bf x},t)$ in terms of the 
new operators in the form 
$$\varphi ({\bf x},t)={\sum_{\bf k}}[A_{\bf k} {\Psi}_{\bf k}(x)+ 
{A^{\dagger}}_{\bf k}{\Psi^{*}}_{\bf k}(x)]\eqno (13)$$
\vskip 0,2 cm
with $\Psi_{\bf k}(x)$ and $\Psi_{\bf k}^{*}(x)$ solutions of eq. (6). 
We can make the separation 
$$\Psi_{\bf k}(x)=h({\bf k},t)\exp {i{\bf k}.{\bf x}}\eqno (14)$$
\vskip 0,2 cm
with
$$h({\bf k},t)={1\over {{{(La(t))}^{3/2}}(2W({\bf k},t))^{1/2}}}
[{\alpha({\bf k},t)}^{*}{e^{-i{\int_{t_{0}}}^{t}Wd{t^{\prime}}}} + 
{\beta({\bf k},t)}^{*}{e^{i{\int_{t_{0}}}^{t}Wd{t^{\prime}}}}]$$
\vskip 0,2 cm
Replacing eq. (14) in eq. (6) we have 
$$(\alpha {e^{i\int Wd{t^{\prime}}}} + \beta {e^{-i\int Wd{t^{\prime}}}})M - 
{{\dot W}\over {W}}({\dot \alpha}e^{i\int Wdt^{\prime}} + 
{\dot \beta}{e^{-i\int Wdt^{\prime}}}) $$
$$ + 2iW({\dot \alpha}
e^{i\int Wdt^{\prime}} - {\dot \beta}{e^{-i\int Wdt^{\prime}}}) + 
{\ddot \alpha}{e^{i\int Wdt^{\prime}}} + {\ddot \beta}
{e^{-i\int Wdt^{\prime}}} = 0\eqno (15)$$
\vskip 0,2 cm
with 
$$M:= -{1\over 2}{({\dot W}/W\dot)} + {1\over 4}{({\dot W}/W)^{2}} - 
{9\over 4}{({\dot a}/a)^{2}} - {3\over 2}{({\dot a}/a\dot)} + 
{{\omega}^{2}} - {W^{2}}\eqno (16)$$

 In the following we will call ${\omega}^{2}={k^{2}/a^{2}}+{m^2}$.

 From eqs (12) and (13), Parker [11] proves that the Bogoliubov coeficients 
satisfy the equation 
$${\dot \beta}=-{\dot \alpha} exp(2i
\int_{t_{0}}^{t}W({\bf k},{t^{\prime}})d{t^{\prime}})\eqno (17)$$

 From the functional form (14), for which the invariance of the 
Klein-Gordon product holds, we obtain:
$$|\alpha|^{2} - |\beta|^{2} = 1\eqno (18)$$

 Using eq.(11) we note that $n_{\bf k}:=|\beta_{\bf k}|^{2}=
<0|{a^{\dagger}}_{\bf k}(t){a_{\bf  k}(t)}|0>$  is  the  mean  value of
 particles created between $t_{1}$ and $t$. Eq. (18) allow us to write 
(as in ref.[11]): 
$$\alpha ({\bf k},t)  =  {e^{-i\gamma_{\alpha}({\bf k},t)}}\cosh \theta ({\bf
k},t)\eqno (19 a)$$
\vskip 0,2 cm
$$\beta ({\bf k},t) =  {e^{i\gamma_{\beta}({\bf  k},t)}}\sinh  \theta  ({\bf
k},t)\eqno (19 b)$$
\vskip 0,2 cm 
$\gamma_{\alpha}$, $\gamma_{\beta}$ and $\theta$ are functions of $\bf k$ and 
$t$ that must satisfy the initial conditions given by eq.(12) and the field 
equation (15).  As we  can  see  from  eq.  (19 a) the number of particles is
only a function of $\theta$, i.e. ; $n_{\bf k}={\sinh}^{2}{\theta}$. 

 Replacing eqs (19) and (17) in eq. (15) and splitting in real and 
imaginary parts, the following system of equations is obtained: 
$$(1 + {\tanh \theta}\cos \Gamma)M + 2W{{\dot \gamma}_{\alpha}} 
= 0\eqno (20a)$$
\vskip 0,2 cm
$$M{\sin \Gamma} + 2W{\dot \theta} = 0\eqno (20 b)$$
\vskip 0,2 cm
with $\Gamma:= {\gamma_{\alpha}} + {\gamma_{\beta}} - 
2\int_{t_{0}}^{t}Wdt^{\prime}$ 

 The solutions to eqs (20) give  us  the  set  of  time  dependent Bogoliubov
transformations  compatible with the Robertson-Walker metric. The  stationary
particle creation is a particular case with $\dot \theta = 0$, therefore 
from eq. (20 b) we have: 
$$M = 0\eqno (21)$$

\bigskip 
3. {\bf VEV of EMT as a thermodynamical object}
\smallskip

 In ref.[4]    the analogy between the Bogoliubov transformation 
which relates  states  at  different  temperature  with  those at
different Cauchy surfaces was shown. Here we will 
exploit this analogy in relation 
with the VEV of EMT. First of all we calculate the classical EMT in curved 
space-time in the usual form (see ref.[2]), i.e. as: 
$$T_{\mu \nu} = {2\over {\sqrt{-g}}} {{\delta S}\over 
{\delta {g^{\mu \nu}}}}$$
\vskip 0,2 cm
we obtain the expression 
$$T_{\mu \nu} = {1\over 2}\{{{\varphi}}_{\mu}, {{\varphi}}_{\nu}\} 
- {1\over 4}
{g_{\mu \nu}}\{{\varphi}^{\sigma},{\varphi}_{\sigma}\} + 
{1\over 4}m^{2} g_{\mu \nu} \{\varphi,\varphi\}\eqno (22)$$
\vskip 0,2 cm
where $\{\ ,\ \}$ is the anticommutator and 
${\varphi_{\mu}}:=\partial_{\mu}\varphi$.

 We can now  calculate  the  EMT  operator  replacing  eq.(13)  in  eq.(22)
obtaining

$${{\hat T}_{\mu\nu}}={\sum _{k k^{\prime}}} \{A_{\bf k}
A_{\bf {k^\prime}}{{\it D}_{\mu\nu}}[{\Psi}_{k}(x),{\Psi}_{k^{\prime}}(x)] 
+ A_{\bf k}{A^{\dagger}}_{\bf k^{\prime}}{{\it D}_{\mu\nu}}[{\Psi}_{\bf k}
(x),{\Psi^\ast}_{\bf    k^{\prime}}(x)]$$ 
$$+ {A^{\dagger}}_{\bf  k}A_{\bf
k^{\prime}}    {\it  D}_{\mu\nu}[{\Psi^\ast}_{\bf    k}(x),{\Psi}_{\bf
k^{\prime}}(x)] + {A^{\dagger}}_{\bf  k}{A^{\dagger}}_{\bf
k^{\prime}}    {\it  D}_{\mu\nu}[{\Psi^\ast}_{\bf k}(x),{\Psi^\ast}_{\bf
k^{\prime}}(x)]$$  
$$+  \{{\bf  k}\leftrightarrow    {\bf  {k^{\prime}}}\}
\}\eqno(23)$$
\vskip 0,2cm
with  $\{{\bf  k}\leftrightarrow    {\bf  {k^{\prime}}}\}$ indicating that 
the last term of eq.(23) is equal to the first changing ${\bf k}$ by 
${\bf k^{\prime}}$.    The differential operator ${\it D}_{\mu\nu}$ is defined
by 
$${{\it D}_{\mu\nu}}[{\varphi},{\psi}]:= {1\over 2}{\partial}_{\mu}{\varphi}
{\partial}_{\nu}{\psi} - {1\over 4}g_{\mu\nu}{{\partial}^{\sigma}}{\varphi}
{{\partial}_{\sigma}}{\psi} + {1\over 4}{m^2} g_{\mu\nu}
{\varphi}{\psi}\eqno (24)$$
\vskip 0,2cm

 We can now calculate the VEV of ${\hat T}_{\mu\nu}$ using the vacuum $|0>$, 
and obtain: 
$$<0|{\hat T}_{\mu\nu}|0>=2{{\sum }_{\bf k}}{\it Re}\{{{\it D}_{\mu\nu}}
[{{\Psi}_{\bf k}}(x), {{{\Psi}^{\ast}}_{\bf k}}(x)]\}\eqno (25)$$

 Replacing now eqs (11) and (13) we also have 
$$\Psi_{\bf k}(x)  = {{\alpha}^{\ast}}\phi_{\bf k}(x) + 
{{\beta}^{\ast}}{{\phi_{-{\bf k}}}^{\ast}}\eqno (26)$$ 
Putting this last equation in eq. (25) yields 
$$<0|{\hat T}_{\mu \nu}|0> = 2{\sum_{\bf k}}{\it Re}\{ (1 + 2|\beta|^{2})
{\it D}_{\mu \nu}[\phi_{\bf k},{{\phi_{\bf k}}^{\ast}}] + 
2{\alpha^{\ast}}{\beta}{\it D}_{\mu \nu}[\phi_{\bf k},\phi_{-{\bf k}}]$$
$$ + {1\over {2}}(|\beta|^2 \dot) E_{\mu \nu}[\phi_{\bf k},
{\phi_{\bf k}}^{\ast}]
 + {1\over {2}}({{\alpha}^{\ast}} \beta \dot) 
E_{\mu \nu}[\phi_{\bf k},\phi_{-{\bf k}}]$$
$$ + {1\over 2}
({\delta_{0 \mu}}{\delta_{0 \nu}} - {1\over {2}}g_{\mu \nu})[(|\dot \alpha|^2 
+ |\dot \beta|^2)|\phi_{\bf k}|^2 + 2{\dot {\alpha^\ast}}
{\dot {\beta^\ast}}{\phi_{-{\bf k}}}{\phi_{\bf k}}]\}\eqno (27)$$     
with 
$$E_{\mu \nu}[\psi,\varphi]:=\psi(\delta_{0\nu}\partial_{\mu}\varphi + 
\delta_{0\mu}\partial_{\nu}\varphi - g_{\mu \nu}\dot \varphi)\eqno (28)$$

  For simplicity we can consider the case in which the particle creation is 
stationary, then $\dot \beta=\dot \alpha=0$, therefore we have 
$$<0|{\hat T}_{\mu \nu}|0> = 2{\sum_{\bf k}}{\it Re}\{ (1 + 2|\beta|^{2})
{\it D}_{\mu \nu}[\phi_{\bf k},{{\phi_{\bf k}}^{\ast}}] + 
2{\alpha^{\ast}}{\beta}{\it D}_{\mu \nu}[\phi_{\bf k},\phi_{-{\bf k}}]
\}\eqno (29)$$

 As we will see the difference between the VEV given by eq. (29) and a 
statistical average is the ``interference" term, for the modes 
$\bf {k}$ and $-\bf {k}$, of eq. (29). 

 A simple way to calculate  the statistical mean value of some physical 
observable is by means of the thermo field dynamics (TFD) formulation 
(see ref. [12]). In this formulation the idea is to take into account the 
interaction between the system and the thermal bath by means of quantum 
fluctuation of the bath (see also ref. [9]), which are called ``tilde" 
fields.  These fields are auxilliary because they are not measurable.  
Those fields
only act in the thermalization of the system modes. The tilde modes are 
associated to tilde creation annhilation operators which act on the vacuum 
of the thermal  bath  $|\tilde {0}>$.    
The  space  of states, in this
formulation is extended in order to include the tilde and the states of the 
system, i.e.: 
$$\{|n,\tilde n>\}=\{|n>\}\otimes \{|\tilde n>\}$$

We will call  the total vacuum 
$\Vert 0>:=|0>\otimes|\tilde 0>$. Moreover we introduce a $T$ parameter in 
order to identify the temperature of the state, i.e.  $|0,T>$ and a Bogoliubov
transformation that relates the vacuum at zero temperature with the one at  
temperature $T$:
$$a_{\bf k}(T)=A_{\bf k}\cosh  \theta({\bf k},T) - {{\tilde A}^{\dagger}}_{\bf
k} \sinh \theta({\bf k},T)\eqno (30)$$

In order to obtain the zero temperature operators,  
$\theta({\bf k},T=0)=0$ is necessary.

 Then we have the creation annhilation bosonic fields $A_{\bf k}$ and 
${A^{\dagger}}_{\bf k}$ which operate in the form 
$$A_{\bf k}\Vert 0> = 0.$$
$${A^{\dagger}}_{\bf k}\Vert 0> = \Vert 1_{\bf k}>, etc.$$ 

with the commutation relation 
$$[A_{\bf k}, {A^{\dagger}}_{{\bf k}^{\prime}}] = 
{\delta}_{{\bf k},{{\bf k}^{\prime}}}$$

 The tilde  operators,  represent  the  quantum  effect  of the reservoir and
satisfy 
$${\tilde A}_{\bf k}\Vert 0> = 0.$$
$${{\tilde A}^{\dagger}}_{\bf k}\Vert 0> = \Vert {\tilde 1}_{\bf k}>, etc.$$
$$[{\tilde A}_{\bf k},{{\tilde A}^{\dagger}}_{{\bf k}^{\prime}}] = 
\delta_{{\bf k},{\bf k}^{\prime}}$$
\vskip 0,2 cm
The thermal operators $a_{\bf k}(T)$, ${a^{\dagger}}_{\bf k}(T)$, 
${\tilde a}_{\bf k}(T)$, ${{\tilde a}^{\dagger}}_{\bf k}(T)$, which satisfy:
$$a_{\bf k}(T)\Vert 0,T>=0$$
$${a^{\dagger}}_{\bf k}(T)\Vert 0,T>=\Vert 1_{\bf k},T>, etc.$$
$$[a_{\bf  k}(T),{a^{\dagger}}_{{\bf  k}^{\prime}}(T)]=\delta_{{\bf   k},{\bf
k}^{\prime}}$$
\vskip 0,2 cm 
and in analogous form the  tilde operators.  The tilde and no 
tilde operators are mutually commutative.  

 With the new Fock space we can calculate the statistical mean value of any 
physical observable,  for  example  if we have the physical operator $\hat A$,
the  mean value can be obtained by means of:
$$<A>=<0,T\Vert {\hat A}\Vert 0,T>$$

 We will calculate  now  the  statistical  mean  value of the energy momentum
operator ${\hat T}_{\mu \nu}$. In orther to do that we can represent the 
operator as a functional of the thermal modes $\phi_{{\bf k} T}$ and 
${\phi^{\ast}}_{{\bf k} T}$ in a way analogous to that in curved space time:   
$${{\hat T}_{\mu\nu}}={\sum _{k k^{\prime}}} \{a_{\bf k}(T)
a_{\bf {k^\prime}}(T){{\it D}_{\mu\nu}}
[{\phi}_{{\bf k}T}(x),{\phi}_{{\bf k}^{\prime}T}(x)] 
+ a_{\bf k}(T){a^{\dagger}}_{\bf k^{\prime}}(T){{\it D}_{\mu\nu}}
[{\phi}_{{\bf k}T}
(x),{\phi^\ast}_{{\bf    k^{\prime}}T}(x)]$$ 
$$+ {a^{\dagger}}_{\bf  k}(T)a_{\bf
k^{\prime}}(T)    {\it  D}_{\mu\nu}[{\phi^\ast}_{{\bf    k}T}(x),{\phi}_{{\bf
k^{\prime}}T}(x)] + {a^{\dagger}}_{{\bf  k}}(T){a^{\dagger}}_{\bf
k^{\prime}}(T)    {\it  D}_{\mu\nu}[{\phi^\ast}_{{\bf k}T}(x),{\phi^\ast}_{{\bf
k^{\prime}}T}(x)]$$  
$$+  \{{\bf  k}\leftrightarrow    {\bf  {k^{\prime}}}\}
\}\eqno(31)$$
\vskip 0,2 cm
It is  easy  to  prove (see ref.  [13]) using the transformation given by eq.
(30) and the inverse that 
$$<0\Vert {{a_{\bf k}}^{\dagger}}(T){a_{\bf k}}(T)\Vert 0>=
<0,T\Vert {{A_{\bf k}}^{\dagger}}A_{\bf k}\Vert 0,T>$$
and in analogous form with $a_{\bf k}(T)a_{\bf k}(T)$ and 
$a_{\bf k}(T){a_{\bf k}}^{\dagger}(T)$. Therefore we can do 
$$<T_{\mu \nu}>=<0\Vert {\hat T}_{\mu \nu}[a_{\bf k}(T),
{a_{\bf k}}^{\dagger}(T),a_{{\bf k}^{\prime}}(T),
{a_{\bf k^\prime}}^{\dagger}(T)]\Vert 0>$$
then the following expression is obtained 
$$<0|{\hat T}_{\mu \nu}|0> = 2{\sum_{\bf k}}{\it Re}\{ (1 + 2n_{\bf k})
{\it D}_{\mu \nu}[\phi_{{\bf k}T},{{\phi_{{\bf k}T}}^{\ast}}] \}\eqno (32)$$
where
$$n_{\bf k}=<0\Vert {a_{\bf k}}^{\dagger}(T)a_{\bf k}(T)\Vert 0>$$
 As we can see from eq.(32) the $\phi_{\bf k}$ modes will be equivalent to 
the thermal ones if the interference term in eq. (29) is null. 

 The difference with the ``genuine thermality" has the same cause as 
that in the 
case of  the  Rindler vs Minkowski observers, analized in ref.  [14].  It is
related with the  fact  that  $<0|a_{\bf k}(t)a_{{\bf k}^{\prime}}(t)|0>\not=0$.
In ref. [14] the thermality is restored assuming that the observation is 
restricted to a finite spacetime region, which is introduced mathematically 
by means of the  wave packets (see also ref. [8]). In our case instead of 
using wave packets we will exploit the fact that the interference term is 
oscillant. First we calculate each component of that term: 
$${\it Re}\{2{\alpha^{\ast}}\beta{\it D}_{00}
[\phi_{\bf k},\phi_{-{\bf k}}]\}=
{1\over 2}{\cosh \theta}{\sinh \theta}{|\phi|^2}\{[{1\over  4}({{\dot W}\over
W}+3H)^{2}$$
$$+{{\omega}^{2}}-{W^{2}}]{\cos \Gamma}-
({{\dot W}\over W}+3H)W{\sin \Gamma}\}\eqno (33)$$ 
$${\it Re}\{ 2{\alpha^{\ast}}\beta{\it D}_{0j}
[\phi_{\bf k},\phi_{-{\bf  k}}]\}=-{\it  Re}\{i{e^{i\gamma}}{{\dot \phi}_{\bf
k}}{\phi_{-{\bf k}}}k_{j}\}\eqno (34)$$
$${\it Re}\{ 2{\alpha^{\ast}}\beta{\it D}_{jj}
[\phi_{\bf k},\phi_{-{\bf  k}}]\}={1\over 2}{\cosh \theta}{\sinh \theta}
{|\phi|^{2}}(2{k^{2}}_{j}-{a^{2}\omega^{2}}){\cos \Gamma}\eqno (35)$$
There are not contributions coming from the term given by eq. (34), because 
when the summation ${\sum_{{\bf k}=-\infty}}^{+\infty}$  is performed, a 
self cancellation is produced.

 The terms given by eqs (33) and (35), when the limit to the continuum is 
taken  $(\sum_{\bf    k}    \to  {(L/2\pi)}^{3}\int  d^{3}{\bf  k})$,  are
proportional to the integrals:
$$I_{00}={{\int_{0}}^{\infty}}{k^2}dk\{[{1\over 4}{({{\dot W}\over W}+3H)^2}+
{\omega^2}-{W^2}]{\cos \Gamma}-({{\dot W}\over W}+3H)W{\sin \Gamma}\}
{\cosh \theta}{\sinh \theta}\eqno (36)$$
$$I_{jj}={{\int_{0}}^{\infty}}{k^2}dk(2{{k_{j}}^2}-{{a^2}{\omega^2}})
{\cos \Gamma}{\cosh \theta}{\sinh \theta}\eqno (37)$$
\vskip 0,2 cm
(where we used ${d^3}{\bf k}={4\pi}{k^2}dk$, and $\Gamma$ as in eqs (20)).
The particle creation is not modified when $\gamma$ changes, as 
we can see from eq. (19). Then we can propose a $\gamma$ phase with the 
functional form 
$$\gamma=\mu k\eqno (38)$$
If we  take  $\mu \to \infty$ and introduce an ultraviolet cutoff (which is
less restritive than the wave packet and can be justified because $k/a$ is the 
physical momentum associated to the $k$ mode, as is shown in ref.[11])
 in the integrals  (36)  and  (37)  we  can  then  apply the Riemann-Lebesgue
theorem,  which is as  follows:    ``If  $f$  is  a  real  function  absolutly
integrable in the interval $[a,b]$, then ${{\lim}_{\gamma \to \infty}}
{\int_{a}^b} f(t) \sin (\gamma t + \delta) dt = 0$ with $\delta$ a real 
constant". Clearly this is also true when we have ``$\cos$" instead of 
``$\sin$".
By means of this theorem  we  can  eliminate  the  interference  term in eq.
(29), and it results equivalent to 
the statistical mean value given by eq.  (32). 
\bigskip
4. {\bf Thermal spectrum condition}

 In order to obtain thermal distribution for the $\bf k$ modes, we need to apply
an extremun condition to a thermodinamical potential.  This potential will be
a function of the energy and the entropy of the bosonic gas in 
curved space-time.  The energy
can be calculated by means of the metric hamiltonian $\hat H$ in the form 
$$E=<0|{\hat H}|0>$$
where
$${\hat H}=\int {a^3}{d^3}x{\hat T}_{00}$$
in the discretized formulation is $\int {d^3}x=L^3$. For eq.(29) when the 
oscillatory term is dropped, we have:
$$E=\sum_{\bf k}({1\over 2}+n_{\bf k}){\epsilon_{\bf k}}\eqno (39)$$
with
$${{\epsilon}_{\bf k}}={1\over {2W}}[{1\over 4}{({{\dot W}\over W}+3H)^2}+
{W^2}+{\omega^2}]\eqno (40)$$
the energy by mode. 
 As the energy for the system studied has the form of a set of decupled quantum 
oscilators, it  can  be  considered  as  a  bosonic  gas  in  flat space time,
therefore we can use the expression for the entropy used in Minkowski 
space. In order to do that we can write the operator (see ref.[12]):
$${\hat K}=-\sum_{k} \{{a_{k}^\dagger}(t)a_{k}(t) \log {{\sinh}^2}\theta 
- a_{k}(t)a_{k}^{\dagger}(t){{\cosh}^2}\theta\}\eqno (41)$$
The expectation value of the operator given by eq.(41) is the entropy of a 
bosonic gas [15]:
$$K=<0|{\hat K}|0>=-\sum_{k}\{{n_{k}}\log{n_{k}}-(1+{n_{k}})
\log(1+{n_{k}})\}\eqno (42)$$
In a  way analogous to ref.[12] we will get an 
extremun, with respect to the Bogoliubov 
angle $\theta$, of the thermodinamical potential  
    
$$\Lambda = -T{ K} + E$$
where  $T$ is  a  Lagrangian  multiplier  that  can  be  interpreted  as  the
temperature. Then performing the variation 
$$\delta {\Lambda}/\delta \theta = 0$$
the following eq. is obtained 
$$T\log {{\sinh^{2} \theta}\over {\cosh^{2} \theta}} + \epsilon =0.$$
therefore we get a planckian spectrum 
$${n_{\bf k}}={1\over {e^{{\epsilon_{\bf k}}\over T} - 1}}\eqno (43)$$
(whit units so that $\hbar=c={k_B}=1$)  
\vskip 0,2cm
 It is interesting to note that we can eliminate the interference term 
in the hamiltonian, without using the phase condition given by eq. (38), 
by means of the criterium  
known as hamiltonian diagonalization. As we can see from eq. (33) 
that condition is 
$${{\dot W}\over W}=-3H\eqno (44a)$$
$${W^2}={\omega^2}\eqno (44b)$$ 
 This condition is, moreover, compatible with the dynamic equation for 
$W$, given by eq. (21). When  at the observation time holds 
eq.   (44),    $D_{00}[\phi_{\bf  k},\phi_{-{\bf  k}}]=0$ 
is assured. Of course the condition (44) has a meaning when the calculation 
is ``in-out", with the parameter ``t" fixed at the observation time.  Then we 
can think of condition (44) as Cauchy data on $W$ and $\dot W$.    
So when 
condition (44) is  satisfied  the  energy  by  mode  is (as we can see from eq.
(40)).
$$\epsilon_{\bf k}=\omega \eqno (45)$$
\vskip 0,2 cm
In particular if we suppose that the rest energy can be dropped against  the 
kinetic one, i. e. ${{k^2}/{a^2}}>>m^2$, we can do 
$\epsilon_{\bf k}\simeq {k/a}$, and obtain
$$n_{\bf k}={1\over {e^{k\over {aT}}-1}}\eqno (46)$$

 From eq.(39)  going  to  the  continuum  set of modes, the particle creation
contribution to the energy density is given by 
$$\rho={1\over {{2{\pi}^{2}}a^{4}}}{{\int_{0}}^{\infty}} k^{3} 
n_{k} dk\eqno (47)$$
\vskip 0,2 cm
 Replacing eq.(46) in eq.(47) and performing the integral, we have 
$$\rho={1\over {2{\pi}^{2}}}{T^{4}}{{\int_{0}}^{\infty}}
{{{k^{\prime}}^{3}}\over {e^{k^{\prime}}-1}}dk^{\prime} = 
{{{\pi}^{2}}\over {30}}T^{4}\eqno (48)$$
\vskip 0,2 cm
which is the expression obtained from the standard phenomenological radiation 
dominated Friedmann cosmology [16].
\bigskip       
  
5.{\bf Conclusions}

 The VEV of the EMT is different to the statistical mean value, even for the 
stationary particle creation. In the last case the difference is an 
interference term between the modes ${\bf k}$ and ${-{\bf k}}$, which can 
be eliminated by a condition on the phase of the Bogoliubov transformation. 
At the 
hamiltonian level the coincidence with the statistical value is satisfied 
by means of the criterion known as diagonalization of the 
hamiltonian condition (that for 
scalar field is identical to the condition of minimization of the 
energy ).  By means of
this condition also  the mean value $<T_{00}>$, defined by 
Utiyama-De Witt [1], and the VEV of the EMT are coincident (see eq.(3)). 

 The coincidence  between  the VEV and the statistical mean value, is another
view of the analogy between the Bogoliubov transformation that 
connects the different vacua in 
curved space-time and the one that relates vacua to different 
temperature in TFD.  
 
 From the  extremal  conditions  on  the  thermodynamical  potential,  we 
obtained a thermal spectrum. The fact that particle creation turns 
to be thermal avoids the existence of 
infinities coming from this effect, in the source of the Einstein equations. 
This is true also when the hamiltonian diagonalization conditions are used. 
In  general  those  conditions  produce  inconsistences  with  the  adiabatic
regularization [4], [17]. All of that suggested the utilization of the 
usual covariant  
regularization tecniques, in order to eliminate the infinities coming 
from  the  vacuum  polarization  term,   together  with  some  
thermodynamic
conditions that ``regularized" the creation particle term. That approach 
allowed us to take into account the back-reaction effect in an easy way, 
without using numerical calculation (see ref.[5]).      
      
\bigskip
{\bf ACKNOWLEDGEMENTS} 

 This work  was  supported  by  the  European  Community  DGXII  and  by  the
Departamento de F\'{\i}sica de la Facultad de Ciencias Exactas y Nat. de 
Buenos Aires. 
\bigskip

{\bf REFERENCES}
\vskip 0,2cm
\item{$[1]$} R.Utiyama  and  B.S.De  Witt;    J.    of Math.  Phys., {\bf 3},
$N^{o}$4 (1962), pp 608-618. 
\item{$[2]$} B.S.De Witt; Phys. Rep. 19, $N^{o}$ 6 (1975), 295-357.   
\item{$[3]$} N.D.Birrell and P.C.W.Davies;  ``Quantum Fields in Curved Space" 
(Cambridge University Press, Cambridge, England, 1982). 
\item{$[4]$} M.Castagnino  and  C.Laciana;    J.  Math.  Phys.  {\bf 29} (2),
(1988), pp 460-475.  
\item{$[5]$} C.E.Laciana; Astrophys. Space Sci., {\bf 229}, $N^{o}$ 2, 
(1995), pp. 173-184. 
\item{$[6]$}   J.D.Barrow,  ``Handbook  of  Astronomy,    Astrophysics    and
Geophysics" Volume II, Galaxies and Cosmology; Edited by V.M.Canuto and 
B.G.Elmegreen, Gordon and Breach Science Publishers, (1983). 
\item{$[7]$} S.W.Hawking; Commun. Math. Phys. {\bf 43}, 199-220 (1975). 
\item{$[8]$} S.Takagi;  Prog.  of Theor.  Phys.  Suppl.  $N^{o}$ 88, 1986, pp
1-142.
\item{$[9]$} H.Umezawa, H.Matsumoto, and M.Tachiki; ``Thermo Field 
Dynamics and Condensed States" (1982) (North-Holland, Amsterdam).
\item{$[10]$} C.E.Laciana; Gen. Rel. and Grav., Vol. {\bf 26}, $N^{o}$ 4, 
(1994), pp 363-378.
\item{$[11]$} L.Parker;  Phys.Rev., Vol.183, $N^{o}$ 5, (1969), pp 1057-1068. 
\item{$[12]$} Y.Takahashi and  H.Umezawa,  (1975).   Collective Phenomena {\bf
2}, 55.  
\item{$[13]$} C.E.Laciana;  ``Particle creation amplification in curved space 
due to thermal effects", Preprint hep-th/9508051.
\item{$[14]$} T.Mishima;  Prog.    of  Theor.    Phys., Vol.  82, $N^{o}$ 5,
(1989), pp 930-944.
\item{$[15]$} L.D.Landau and E.M.Lifshitz, ``Statistical Physics", Vol.5. 
Pergamon Press (2nd Impression 1970), eq. 54.6, pag. 147.  
\item{$[16]$} A.Linde;  ``Particle Physics and Inflationary Cosmology", (1990)
(Harwood Academic Publishers, NY), 91. 
\item{$[17]$} M.Castagnino and C.Laciana; Prog. of Theor. Phys., Vol. 84, 
$N^{o}$ 4, (1990), pp 595-615.

\bye